\documentclass[floatfix,aps,showpacs,twocolumn,nofootinbib,preprintnumbers]{revtex4}
\usepackage{graphicx}

\begin{document}

\preprint{CERN-TH/2005-213}

\title{Theoretical Uncertainties in Inflationary Predictions}
\author{William H.\ Kinney} \email{whkinney@buffalo.edu}
\affiliation{Dept. of Physics, University at Buffalo,
        the State University of New York, Buffalo, NY 14260-1500}
\author{Antonio Riotto} \email{antonio.riotto@pd.infn.it}
\affiliation{CERN, Theory Division, CH-1211 Geneva 23, Switzerland}
\date{\today}

\begin{abstract}
\noindent 
With present and future observations becoming of higher and higher quality,
it is  timely and necessary to investigate the most significant theoretical
uncertainties in the predictions of inflation. We show that our ignorance
of the entire history of the Universe, including the 
physics of reheating after
inflation, translates to considerable errors in observationally
relevant parameters. Using the inflationary flow formalism, we estimate that 
for a spectral index $n$ and tensor/scalar ratio $r$ in the region favored by 
current observational constraints, the theoretical errors are of order
 $\Delta n / \left\vert n - 1\right\vert \sim 0.1 - 1$ and $\Delta r /r \sim 0.1 - 1$.
These errors represent the dominant theoretical 
uncertainties in the predictions
of inflation, and are  generically of the order of or larger 
than the projected uncertainties 
in future precision measurements of the Cosmic Microwave Background.
We also show that the lowest-order 
classification of models into small field, large field, and hybrid breaks 
down when higher order corrections to the dynamics are included. Models
can flow from one region to another.
\end{abstract}


\maketitle

\section{Introduction}

Inflation \cite{lrreview} has become the dominant paradigm for
understanding the initial conditions for structure formation and for
Cosmic Microwave Background (CMB) anisotropies. In the inflationary
picture, primordial density and gravitational-wave fluctuations are created
from quantum fluctuations, ``redshifted'' beyond the horizon during an
early period of superluminal expansion of the universe, then
``frozen'' \cite{Starobinsky:1979ty,muk81,bardeen83}. Perturbations at the surface of
last scattering are observable as temperature anisotropies in the CMB,
as first detected by the Cosmic Background Explorer 
satellite \cite{bennett96,gorski96}. The latest and most impressive
confirmation of the inflationary paradigm has been recently provided
by data from the Wilkinson Microwave Anisotropy Probe (WMAP)
satellite, which marks the beginning of the precision era of CMB
measurements in space \cite{wmap1}. The WMAP collaboration has
produced a full-sky map of the angular variations of the CMB to
unprecedented accuracy. WMAP data support the inflationary mechanism
as the mechanism for the generation of super-horizon curvature
fluctuations. 

The CMB contains a wealth of information
about the properties of the spectrum of  primeval density perturbations and
present data already allow to extract relevant informations about 
the parameters of single-field models of inflation
\cite{ex}, {\it i.e.} models whose inflation is driven by 
one scalar field, the inflaton. The following parameters have been identified
as important for accurately computing the
expected anisotropy and for discriminating among different inflationary
models: the power-law indices of the scalar and tensor perturbations
$n$ and $n_T$ respectively and the 
tensor-to-scalar amplitude ratio $r=16 (P_T/P_{\cal R})$. Present
data are consistent with a scale-invariant spectrum of
scalar perturbation ($n=1$) and with an amount of tensor perturbations
such that   
$r\lesssim 0.5$ \cite{ex}. However, future 
CMB experiments will allow an accurate  determination of the properties
the scalar spectrum. The satellite-borne experiment Planck
\cite{planck}, the proposed high-resolution version of CMBpol \cite{CMBpol}
and a polarized bolometer array on the South Pole Telescope \cite{south}
will allow a determination of the spectral index $n$ with a standard
deviation of about 0.007, 0.003 and 0.01, respectively \cite{knox}.
At the same time a  positive detection of the tensor modes
through the $B$-mode of CMB polarization (once foregrounds due to
gravitational lensing from local sources have been properly treated)
requires $r \gtrsim 10^{-3}$
\cite{gravex}. While this limit is below the expected sensitivity, a tensor
to scalar ratio of $r\sim 0.01$ is well within the reach of presently feasible
CMB observations. The proposed Big Bang Observer satellite has the potential to probe
$r \sim 0.001$ \cite{BBO}.

With present and future observations reaching a higher and higher quality,
it becomes  timely and necessary to investigate the most significant
uncertainties on the theoretical side as far as inflationary predictions are
concerned. In this paper 
we study the impact of our ignorance about the
precise location on the inflationary potential corresponding
to the observed perturbations. This is quantified 
by the number of {\it e}-foldings $N$ before the end of inflation at which
our present Hubble scale equalled the Hubble scale during inflation, the
so-called epoch of horizon-crossing. Indeed, the determination
of the number of {\it e}-foldings requires the knowledge
of the entire history of the Universe.
The expression for $N$ can be written as

\begin{equation}
N\simeq 60+\frac{1}{6}{\rm ln}\,\left(-n_T\right)+\frac{1}{3}\,{\rm ln}
\left(T_{\rm RH}/10^{16}\,{\rm GeV}\right)-\frac{1}{3}\,{\rm ln}\,\gamma\, ,
\end{equation}
where $T_{\rm RH}$ is the reheating temperature, $\gamma$ is the ratio of the
entropy per comoving volume today to that after reheating and quantifies
any post-inflation entropy production and we have assumed that there is no
significant drop in energy density during the last stages of inflation. The 
main uncertainties in the determination of the number of {\it e}-foldings
are  caused by our ignorance about the last two terms. 
The reheating temperature $T_{\rm RH}$ 
after inflation may vary from the Grand Unified Theory
 scale $\sim 10^{16}$ GeV to
1 MeV, the scale at which nucleosynthesis takes place. In this range, the 
corresponding shift of $N$ is about 14. Furthermore, long-lived massive
particles of mass of the order of the weak scale are ubiquitous
in string-inspired models (they are generically dubbed moduli) 
and  may dominate the energy density of the
Universe after reheating,  leading to a prolonged
matter-dominated epoch followed by a large amount of entropy release
at the time of moduli decay \cite{decarlos}, $\gamma\gg 1$ . 
The corresponding shift in the number of {\it foldings} can be as large
as 10. One can even envisage extreme situations where the reduction of the
energy scale during  inflation is so 
significant  that the shift in $N$ is as large as 70 \cite{ll}. 

Given the
fact that the values of the  inflationary observables $n$, $n_T$ and $r$
are evaluated at the  value of $N$ corresponding to the moment when the
present Hubble scale crossed outside the horizon during inflation and 
that such a value is affected by a non-negligible uncertainty, we
immediately   conclude that the predictions 
of the inflationary observables are affected by  unavoidable  theoretical 
errors. How can we quantify them? Are they larger or smaller 
than the expected accuracy
of forthcoming experiments?
We propose 
to  use the method of ``flow'' to gain 
some insight.  The flow equations provide   the 
derivatives of the inflationary observables with respect to the number 
of {\it e}-foldings 
as a function of the observable themselves at any order in the so-called  
slow-roll parameters \cite{kinneyflow,hoffman,sc}. For instance, 
to lowest order in slow roll

\begin{eqnarray}
\label{flow1}
\frac{d r}{d N} &=& r \left[(n - 1) + \frac{r}{8}\right]\, ,\nonumber\\
\frac{d n}{d N} &=& -\frac{5}{16}\, r (n - 1) - \frac{3}{32} r^2 + 2 \xi^2\, ,
\end{eqnarray}
where $\xi^2\equiv (m_{\rm Pl}^4/16\pi^2)\left(H^\prime 
H^{\prime\prime\prime}/H^2\right)$, $m_{\rm Pl}$ being the Planck scale,
$H$ the Hubble rate during inflation and primes indicate differentiation
with respect to the inflaton field. From this set of equations we can easily
quantify  -- within any given single-field model of inflation --
how our ignorance on the precise value of the 
number of {\it e}-foldings, quantified by 
a  shift in the number of {\it e}-foldings $\Delta N$, is reflected in 
the predicted value of the observable quantities. The expected uncertainties are model-dependent, but roughly speaking,
we expect a theoretical error of the magnitude $\Delta r\sim (dr/dN)_*\Delta N$
and $\Delta n\sim (d n/dN)_*\Delta N$, where the derivatives are evaluated
at a reference  number of {\it e}-folding, {\it e.g.} at $N_*\sim 60$ 
corresponding to $T_{\rm RH}\sim 10^{16}$ GeV and $\gamma\sim 1$ and
$\Delta N=(N-N_*)$. 
We  conclude that -- generically -- the error
in the tensor-to-scalar ratio is of order $\Delta r/r\sim 0.1 - 1$, and the error
in the spectral index 
is of order $\Delta n / \left\vert n - 1\right\vert\sim 0.1 - 1$ (for $\Delta N\sim 14$). These errors 
are of the order of or larger than the accuracy expected from future experiments. 

The paper is organized as follows. In Sec.\ II we discuss single-field
inflation and the relevant observables in more detail. In Sec.\ III we discuss
the inflationary model space, and in Sec.\ IV we describe the flow technique
to quantify the theoretical errors in  the inflationary predictions. 
In Sec.\ V we present our results and offer some analytical explanations
Finally, in Sec.\ VI we present our conclusions.

\section{Single-field inflation and the inflationary observables}

In this section we briefly review scalar field models of inflationary
cosmology, and explain how we relate model parameters to observable quantities.
Inflation, in its most general sense, can be defined to be a period of
accelerating cosmological expansion during which the universe evolves toward
homogeneity and flatness. This acceleration is typically a result of the
universe being dominated by vacuum energy, with an equation of state $p \simeq
-\rho$. Within this broad framework, many specific models for inflation have
been proposed. We limit ourselves here to models with ``normal'' gravity ({\em
i.e.,} general relativity) and a single order parameter for the vacuum,
described by a slowly rolling scalar field $\phi$, the inflaton. 

A scalar field in a cosmological background evolves with an equation of motion
\begin{equation}
\ddot\phi + 3 H \dot\phi + V'\left(\phi\right) = 0.
\label{eqequationofmotion}
\end{equation}
The evolution of the scale factor is given by the scalar field dominated FRW
equation,
\begin{eqnarray}
H^2 & &= {8 \pi \over 3 m_{\rm Pl}^2} \left[{1 \over 2} \dot\phi^2 +
V\left(\phi\right)\right],\cr
\left(\ddot a \over a\right) &&= {8 \pi \over 3 m_{\rm Pl}^2}
\left[V\left(\phi\right) - \dot\phi^2\right].
\label{eqbackground}
\end{eqnarray}
We have assumed a flat Friedmann-Robertson-Walker metric,
\begin{equation}
g_{\mu \nu} =  {\rm diag}(1, -a^2, -a^2 -a^2),
\end{equation}
where $a^2(t)$ is the scale factor of the universe.  {\em Inflation} is defined
to be a period of accelerated expansion, $\ddot a > 0$.  A powerful way of
describing the dynamics of a scalar field-dominated cosmology is to express the
Hubble parameter as a function of the field $\phi$, $H = H(\phi)$, which is
consistent provided $\phi$ is monotonic in time. The equations of motion
become \cite{grishchuk88,muslimov90,salopek90,lidsey95}:
\begin{eqnarray} 
& &\dot\phi =  -{m_{\rm Pl}^2 \over 4 \pi} H'(\phi),\cr
& & \left[H'(\phi)\right]^2 - {12 \pi \over m_{\rm Pl}^2}
H^2(\phi) = - {32 \pi^2 \over m_{\rm Pl}^4}
V(\phi).\label{eqbasichjequations}
\end{eqnarray}
These are completely equivalent to the second-order equation of motion in Eq.\
(\ref{eqequationofmotion}). The second of the above equations is
referred to as the {\it Hamilton-Jacobi} equation, and can be written
in the useful form
\begin{equation} 
H^2(\phi) \left[1 - {1\over 3}
\epsilon(\phi)\right] =  \left({8 \pi \over 3 m_{\rm Pl}^2}\right)
 V(\phi),\label{eqhubblehamiltonjacobi}
\end{equation}
where $\epsilon$ is defined to be
\begin{equation}
\epsilon \equiv {m_{\rm Pl}^2 \over 4 \pi} \left({H'(\phi) \over
 H(\phi)}\right)^2.\label{eqdefofepsilon}
\end{equation}
The physical meaning of $\epsilon$ can be seen by expressing Eq.\
(\ref{eqbackground}) as
\begin{equation}
\left({\ddot a \over a}\right) = H^2 (\phi) \left[1 -
 \epsilon(\phi)\right],
\end{equation}
so that the condition for inflation $(\ddot a / a) > 0$ is given by
$\epsilon < 1$. The scale factor is given by
\begin{equation}
a \propto e^{N} = \exp\left[\int_{t_0}^{t}{H\,dt}\right],
\end{equation}
where the number of {\it e}-folds $N$ is
\begin{equation}
N \equiv \int_{t}^{t_e}{H\,dt} = \int_{\phi}^{\phi_e}{{H \over
\dot\phi}\,d\phi} = {2 \sqrt{\pi} \over m_{\rm Pl}}
\int_{\phi_e}^{\phi}{d\phi \over
\sqrt{\epsilon(\phi)}}.\label{eqdefofN}
\end{equation}
whose value has been discussed in the Introduction. 

We will frequently work within the context of the {\em slow roll} approximation
which is the assumption that the evolution of the
field is dominated by drag from the cosmological expansion, so that $\ddot\phi
\simeq 0$ and
\begin{equation}
\dot \phi \simeq -{V' \over 3 H}.
\end{equation}
The equation of state of the scalar field is dominated by the potential,
so that $p \simeq -\rho$, and the expansion rate is approximately
\begin{equation}
H \simeq \sqrt{{8 \pi \over 3 m_{\rm Pl}^2} V\left(\phi\right)}.
\label{eqhslowroll}
\end{equation}
The slow roll approximation is consistent if both the slope and curvature of
the potential are small, $V',\ V'' \ll V$. In this case the parameter
$\epsilon$ can be expressed in terms of the potential as
\begin{equation}
\epsilon \equiv {m_{\rm Pl}^2 \over 4 \pi} \left({H'\left(\phi\right) \over
H\left(\phi\right)}\right)^2 \simeq {m_{\rm Pl}^2 \over 16 \pi}
\left({V'\left(\phi\right) \over V\left(\phi\right)}\right)^2.
\end{equation}
We will also define a second ``slow roll parameter'' $\eta$ by:
\begin{eqnarray}
\eta\left(\phi\right) &\equiv& {m_{\rm Pl}^2 \over 4 \pi} 
\left({H''\left(\phi\right)
\over H\left(\phi\right)}\right)\cr
&\simeq& {m_{\rm Pl}^2 \over 8 \pi}
\left[{V''\left(\phi\right) \over V\left(\phi\right)} - {1 \over 2}
\left({V'\left(\phi\right) \over V\left(\phi\right)}\right)^2\right].
\end{eqnarray}
Slow roll is then a consistent approximation for $\epsilon,\ \eta \ll 1$. 

Inflation models not only explain the large-scale homogeneity of the universe,
but also provide a mechanism for explaining the observed level of {\em
inhomogeneity} as well. During inflation, quantum fluctuations on small scales
are quickly redshifted to scales much larger than the horizon size, where they
are ``frozen'' as perturbations in the background metric. 
The metric perturbations
created during inflation are of two types: scalar, or {\it curvature}
perturbations, which couple to the stress-energy of matter in the universe and
form the ``seeds'' for structure formation, and tensor, or gravitational wave
perturbations, which do not couple to matter.  Both scalar and tensor
perturbations contribute to CMB anisotropy. Scalar fluctuations can also be
interpreted as fluctuations in the density of the matter in the
universe. Scalar fluctuations can be quantitatively characterized by
the comoving curvature perturbation $P_{\cal R}$. As long as the
equation of state $\epsilon$ is slowly varying,  the curvature
perturbation can be shown to be 
\cite{lrreview}
\begin{equation}
P_{\cal R}^{1/2}\left(k\right) = \left({H^2 \over 2 \pi \dot \phi}\right)_{k = a H} =    \left [{H \over m_{\rm Pl} }
{1 \over \sqrt{\pi \epsilon}}\right]_{k = a H}.
\end{equation}
The fluctuation power spectrum is in general a function of wavenumber $k$, and
is evaluated when a given mode crosses outside the horizon during inflation, $k
= a H$. Outside the horizon, modes do not evolve, so the amplitude of the mode
when it crosses back {\em inside} the horizon during a later radiation- or
matter-dominated epoch is just its value when it left the horizon during
inflation. 
Instead of specifying the fluctuation amplitude directly as a function of $k$,
it is convenient to specify it as a function of the number of {\it e}-folds $N$
before the end of inflation at which a mode crossed outside the horizon.

The {\em spectral index} $n$ for $P_{\cal R}$ is defined by
\begin{equation}
n - 1 \equiv {d\ln P_{\cal R} \over d\ln k},
\end{equation}
so that a scale-invariant spectrum, in which modes have constant amplitude at
horizon crossing, is characterized by $n = 1$. 

The power spectrum of tensor fluctuation modes is given
by \cite{lrreview}
\begin{equation}
P_{T}^{1/2}\left(k_N\right) = \left[\frac{4 H}{m_{\rm Pl} \sqrt{\pi}}
\right]_{N}.
\end{equation}
The ratio of tensor-to-scalar modes is then
\begin{equation}
{P_{T} \over P_{\cal R}} = 16 \epsilon,
\end{equation}
so that tensor modes are negligible for $\epsilon \ll 1$.

\section{The inflationary model space}
\label{seczoology}

To summarize the results of the previous section, inflation generates scalar
(density) and tensor (gravity wave) fluctuations which are generally well
approximated by power laws:
\begin{equation}
P_{\cal R}\left(k\right) \propto k^{n - 1}; \qquad
P_{T}\left(k\right) \propto k^{n_{T}}.
\end{equation}
In the limit of slow roll, the spectral indices $n$ and $n_{T}$ vary slowly 
or not at all with scale.
We can write the spectral indices $n$ and $n_{T}$ to lowest order in terms
of the slow roll parameters $\epsilon$ and $\eta$ as:
\begin{eqnarray}
n \simeq&& 1 - 4 \epsilon + 2 \eta,\cr
n_{T} \simeq&& - 2 \epsilon.
\end{eqnarray}
The tensor/scalar ratio is frequently expressed as a quantity $r$ which is 
conventionally normalized as
\begin{equation}
r \equiv 16 \epsilon = {P_{\rm T} \over P_{\cal R}}
\end{equation}
The tensor spectral index is {\em not} an independent parameter, but
is proportional to the tensor/scalar ratio, given to lowest order in
slow roll by
\begin{equation}
n_{T} \simeq - 2 \epsilon = - {r \over 8}.
\end{equation}
This is known as the {\it consistency relation} for inflation.
 A given inflation model can therefore be
described to lowest order in slow roll by three independent parameters,
$P_{\cal R}$, $P_{T}$, and $n$. If we wish to include higher-order effects, we
have a fourth parameter 
describing the running of the scalar spectral index, $d
n / d\ln{k}$.

Calculating the CMB fluctuations from a particular inflationary model reduces
to the following basic steps: (1) from the potential, calculate $\epsilon$ and
$\eta$. (2) From $\epsilon$, calculate $N$ as a function of the field $\phi$.
(3) Invert $N\left(\phi\right)$ to find $\phi_N$. (4) Calculate $P_{\cal R}$,
$n$, and $P_T$ as functions of $\phi$, and evaluate them at $\phi =
\phi_N$. For the remainder of the paper, all parameters are assumed to be
evaluated at $\phi = \phi_N$.  

Even restricting ourselves to a simple single-field inflation scenario, the
number of models available to choose from is large \cite{lrreview}.  It is
convenient to define a general classification scheme, or ``zoology'' for models
of inflation. We divide models into three general types: {\it large-field},
{\it small-field}, and {\it hybrid}, with a fourth classification, {\it linear}
models, serving as a boundary between large- and small-field. A generic
single-field potential can be characterized by two independent mass scales: a
``height'' $\Lambda^4$, corresponding to the vacuum energy density during
inflation, and a ``width'' $\mu$, corresponding to the change in the field
value $\Delta \phi$ during inflation:
\begin{equation}
V\left(\phi\right) = \Lambda^4 f\left({\phi \over \mu}\right).
\end{equation}
Different models have different forms for the function $f$. The height
$\Lambda$ is fixed by normalization, so the only free parameter is the width
$\mu$.

With the normalization fixed, the relevant parameter space for distinguishing
between inflation models to lowest order in slow roll is then the $r\,-\,n$
plane.  (To next order in slow-roll parameters, one must introduce the running
of $n$.)  Different classes of models are distinguished by the value of the
second derivative of the potential, or, equivalently, by the relationship
between the values of the slow-roll parameters $\epsilon$ and
$\eta$. 
Each class of models has a different relationship between $r$ and $n$. For a
more detailed discussion of these relations, the reader is referred to Refs.\
\cite{dodelson97,kinney98a}.  

First order in $\epsilon$ and $\eta$ is sufficiently accurate for the purposes
of this Section, and for the remainder of this Section we will only work to
first order. The generalization to higher order in slow roll will be discussed 
in the following.

\subsection{Large-field models: $-\epsilon < \eta \leq \epsilon$}

Large-field models have inflaton potentials typical of ``chaotic'' inflation
scenarios \cite{linde83}, in which the scalar field is displaced from the
minimum of the potential by an amount usually of order the Planck mass. Such
models are characterized by $V''\left(\phi\right) > 0$, and $-\epsilon < \eta
\leq \epsilon$. The generic large-field potentials we consider are polynomial
potentials $V\left(\phi\right) = \Lambda^4 \left({\phi / \mu}\right)^p$,
and exponential potentials, $V\left(\phi\right) = \Lambda^4 \exp\left({\phi /
\mu}\right)$. 

For the case of an exponential potential, $V\left(\phi\right)
\propto \exp\left({\phi / \mu}\right)$, the tensor/scalar ratio $r$ is simply
related to the spectral index as
\begin{equation}
r = 8 \left(1 - n\right),
\end{equation}
but the slow roll parameters are constant (there is no dependence 
upon $N$) and therefore no intrinsic errors of the
observables $n$ and $r$ are expected in such a case.

For inflation with a
polynomial potential, $V\left(\phi\right) \propto \phi^p$,  we 
have

\begin{eqnarray}
n-1&=&-\frac{2+p}{2N}\, ,\nonumber\\
r&=&\frac{8p}{2N}=8 \left({p \over p + 2}\right) \left(1 - n\right)\, ,
\end{eqnarray}
so that tensor modes are large for significantly tilted spectra. 
By shifting the number of {\it e}-foldings by $\Delta N$ one therefore expects

\begin{equation}
\label{large}
\frac{\Delta(n-1)}{n-1}=\frac{\Delta r}{r}=-\frac{\Delta N}{N}\, .
\end{equation}
From these relations we deduce that sizeable  correlated theoretical errors should be
expected for those large-field models characterized by  large deviations from 
a flat spectrum and by large values of the tensor-to-scalar 
amplitude ratio. Furthermore these errors increase with the potential
of the polynomial $p$. Of course, these statements are based on relations valid only
at first order in the slow roll parameters. This means that for very large values
of $(n-1)$ and $r$ higher order corrections become relevant and may significantly
alter the simple relations (\ref{large}).

\subsection{Small-field models: $\eta < -\epsilon$}

Small-field models are the type of potentials that arise naturally from
spontaneous symmetry breaking (such as the original models of ``new'' inflation
\cite{linde82,albrecht82}) and from pseudo Nambu-Goldstone modes (natural
inflation \cite{freese90}). The field starts from near an unstable equilibrium
(taken to be at the origin) and rolls down the potential to a stable
minimum. Small-field models are characterized by $V''\left(\phi\right) < 0$ and
$\eta < -\epsilon$. Typically $\epsilon$ (and hence the tensor amplitude) is
close to zero in small-field models. The generic small-field potentials we
consider are of the form $V\left(\phi\right) = \Lambda^4 \left[1 - \left({\phi
/ \mu}\right)^p\right]$, which can be viewed as a lowest-order Taylor expansion
of an arbitrary potential about the origin. The cases $p = 2$ and $p > 2$ have
very different behavior. For $p = 2$, $n-1\simeq -(1/2\pi)(m_{\rm Pl}/\mu)^2$ and there is
no dependence upon the number of {\it e}-foldings. On the other hand

\begin{equation}
r = 8 (1 - n) \exp\left[- 1 - N\left(1 - n\right)\right],
\end{equation}
leading to

\begin{equation}
\frac{\Delta r}{r}=(n-1)\Delta N\, .
\end{equation}
 For $p > 2$, the scalar
spectral index is
\begin{equation}
n \simeq 1 - {2 \over N} \left({p - 1 \over p - 2}\right),
\end{equation}
{\it independent} of $(m_{\rm Pl}/\mu)$. 
Assuming $\mu < m_{\rm Pl}$ results in an upper bound
on $r$ of
\begin{equation}
r < 8 {p \over N \left(p - 2\right)} \left({8 \pi \over N p \left(p -
2\right)}\right)^{p / \left(p - 2\right)}.
\end{equation}
The corresponding theoretical errors read

\begin{eqnarray}
\label{small}
\frac{\Delta(n-1)}{n-1}&=&-\frac{\Delta N}{N}\, ,\nonumber\\
\frac{\Delta r}{r}&\simeq& (n-1)\Delta N\simeq -
\frac{2(p-1)}{p-2}\frac{\Delta N}{N}\, .
\end{eqnarray}
Due to the tiny predicted values of $r$, for small field models one expects 
generically tiny errors in the tensor-to-scalar 
amplitude ratio, but sizeable errors
in the spectral index. 

\subsection{Hybrid models: $0 < \epsilon < \eta$}

The hybrid scenario \cite{linde91,linde94,copeland94} frequently appears in
models which incorporate inflation into supersymmetry. In a typical hybrid
inflation model, the scalar field responsible for inflation evolves toward a
minimum with nonzero vacuum energy. The end of inflation arises as a result of
instability in a second field. Such models are characterized by
$V''\left(\phi\right) > 0$ and $0 < \epsilon < \eta$. We consider generic
potentials for hybrid inflation of the form $V\left(\phi\right) = \Lambda^4
\left[1 + \left({\phi / \mu}\right)^p\right].$ The field value at the end of
inflation is determined by some other physics, so there is a second free
parameter characterizing the models. Because of this extra freedom, hybrid
models fill a broad region in the $r\,-\,n$ plane. For $\left({\phi_N / \mu}\right)\gg
1$ (where $\phi_N$ is the value of the inflaton field when there are
{\it e}-foldings till the end of inflation) one recovers the same results of the large
field models. On the contrary, when 
$\left({\phi_N / \mu}\right)\ll 1$, the dynamics are analogous to small-field models, except that the field is evolving toward, rather than away from, a dynamical fixed point. This distinction is important to the discussion here because near the fixed point the parameters $r$ and $n$ become independent of the number of {\it e}-folds $N$, and the corresponding theoretical uncertainties due to the uncertainty in $N$ vanish.  However, there is an additional degree of freedom not present in other models due to the presence of the additional parameter $\phi_c$. Therefore the theoretical uncertainties in
the predictions of a generic hybrid inflation model are decoupled from
the physics of reheating, and we do not consider such models further here. 
The distinguishing observational feature of many hybrid models is $\eta > 0$ and a {\it
blue} scalar spectral index, $n > 1$. 

Notice that at first order in the slow roll parameters, there is 
no overlap in the $r\,-\,n$ plane between hybrid inflation
and other models. However, as we will explicitly show, this feature is lost
going beyond first order: by changing $N$ models can flow from the hybrid
regions to other model regions; this feature is generic, models can flow from one
region to another. Therefore it is important to distinguish between models
labeled ``hybrid'' in the sense of evolution toward a late-time asymptote and
the {\em region} labeled ``hybrid'' in the zoo plot. The lowest-order correspondence
does not always survive to higher order in slow roll.

\subsection{Linear models: $\eta = - \epsilon$}

Linear models, $V\left(\phi\right) \propto \phi$, live on the boundary between
large-field and small-field models, with $V''\left(\phi\right) = 0$ and $\eta =
- \epsilon$. The spectral index and tensor/scalar ratio are related as:
\begin{equation}
r = {8 \over 3} \left(1 - n\right).
\end{equation}
For linear models, Eq. (\ref{large}) applies. 

This enumeration of models is certainly not exhaustive. There are a number of
single-field models that do not fit well into this scheme, for example
logarithmic potentials $V\left(\phi\right) =V_0\left[1+(C g^2/8\pi)
\ln\left(\phi/\mu\right)\right]$ typical of supersymmetry
\cite{lrreview}, where $C$ counts the degrees of freedom coupled
to the inflaton field and $g$ is a coupling constant.
 For this kind of potentials, one gets $n-1\simeq -(1/N)$ and $r\simeq 
(2 C g^2/\pi^2)(1/N)$ corresponding to 

\begin{equation}
\label{oneloop}
\frac{\Delta(n-1)}{n-1}=\frac{\Delta r}{r}=-\frac{\Delta N}{N}\, .
\end{equation}
Because of the loop-factor suppression, one typically expects 
tiny theoretical errors in $r$, but sizeable uncertainties in $n-1$.

Another example is potentials
with negative powers of the scalar field $V\left(\phi\right) =V_0\left[
1+\alpha\left(m_{\rm Pl}/\phi\right)^p\right]$ 
used in intermediate inflation \cite{barrow93} and dynamical
supersymmetric inflation \cite{kinney97,kinney98}. Both of these cases require
an auxiliary field to end inflation and are more properly categorized as
hybrid models, but fall into the small-field region of the $r\,-\,n$ plane.
The power spectrum is blue being the spectral index  
given by $n-1\simeq 2(p+1/p+2)(2/(N_{\rm tot}-N))$, where
$N_{\rm tot}$ is the total number of {\it e}-foldings; the parameter
$r$ turns out to be proportional to
$(n-1)^{2(p+1)/(p+2)}$. Therefore,

\begin{equation}
\label{dynamical}
\frac{\Delta(n-1)}{n-1}=\frac{p+2}{2(p+1)}\frac{\Delta r}{r}
=-\frac{\Delta N}{N_{\rm tot}-N}\, .
\end{equation}
Uncertainties in the spectral index can be sizeable if 
$N_{\rm tot}$ is close to $N$, but the theoretical errors
in $r$ are expected to be suppressed for small $r$.

\begin{figure}
\includegraphics[width=3.25in]{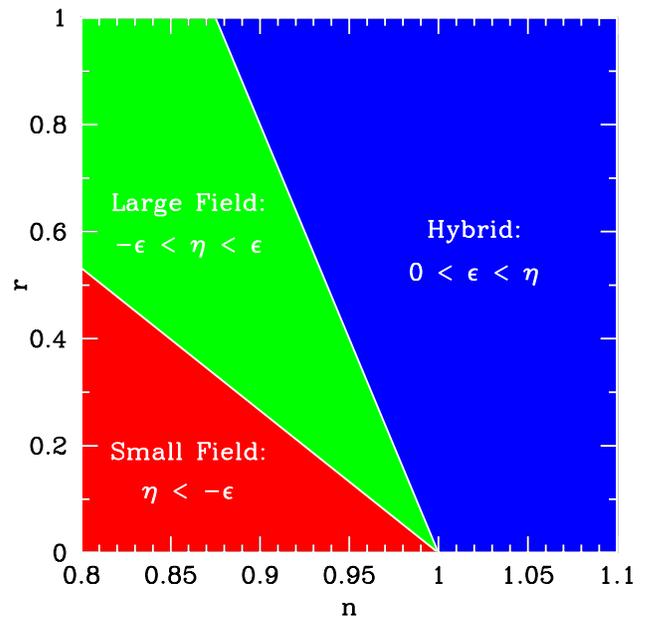}
\caption{\label{fig:regions} Regions in the $r\,-\,n$ plane corresponding to ``large field'', ``small field'', and ``hybrid'' models.}
\end{figure}

\begin{figure}
\includegraphics[width=3.25in]{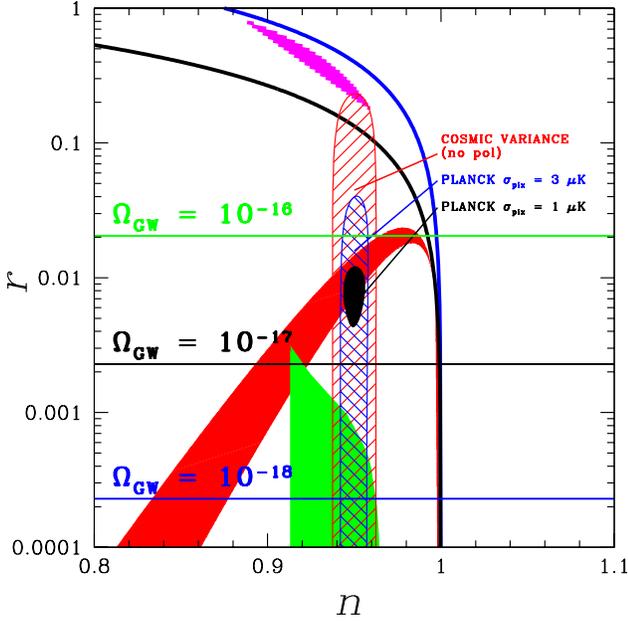}
\caption{\label{fig:modelslog} Predictions of different potentials, including errors due to the uncertainty in $N$, plotted in the region of $n\,-\,\log{r}$ plane favored by current observation. The model curves, from largest $r$ to smallest, are: $V \propto \exp(\phi / \mu)$ (blue, top), $V \propto \phi^p$ (magenta), $V \propto \phi$ (black), $V \propto 1 - (\phi / \mu)^2$ (red), $V \propto 1 - (\phi / \mu)^p\, (p > 2)$ (green, lowest). The horizontal lines labeled with $\Omega_{\rm GW}$ are the expected sensitivities of different proposed configurations of the Big Bang Observer satellite \cite{BBO}. The hatched error bars are the observational uncertainties for: (a) a cosmic-variance limited temperature-only CMB measurement to $\ell = 1500$ (outer, red), the Planck Surveyor satellite (middle, blue), and a hypothetical CMBPol-like experiment with the same angular resolution as Planck but three times better sensitivity (inner, solid black) \cite{kinney98a}. The central value for the error bars shown is arbitrary.}
\end{figure}

The three classes categorized by the relationship between the
slow-roll parameters as $-\epsilon <
\eta \leq \epsilon$ (large-field), $\eta \leq -\epsilon$ (small-field, linear),
and $0 < \epsilon < \eta$ (hybrid), cover the entire $r\,-\,n$ plane and are in
that sense complete (at least at first order in the slow roll parameters) 
Figure \ref{fig:regions}
\cite{dodelson97} shows the $r\,-\,n$ plane divided into regions representing
the large field, small-field and hybrid cases. Figure \ref{fig:modelslog} \cite{kinney98a}
shows a ``zoo plot'' of
the particular potentials considered here plotted on the $n\,-\,\log{r}$ plane, along
with projected errors from forthcoming experiments. For a given choice of potential of the form 
\begin{equation}
V\left(\phi\right) = \Lambda^4 f\left({\phi \over \mu}\right),
\end{equation}
the parameter $\Lambda$ is generally fixed by CMB normalization, leaving the mass scale $\mu$ and the number of {\it e}-folds $N$ as free parameters. For some choices of potential,
for example $V \propto \exp{(\phi / \mu)}$ or $V \propto 1 - (\phi / \mu)^2$, the spectral index $n$ varies as a function of $\mu$. These models therefore appear for fixed $N$ as lines on the zoo plot. The inclusion of the uncertainty in $N$ results in a broadening of the line. For other choices of potential, for example $V \propto 1 - (\phi / \mu)^p$ with $p > 2$, the spectral index is independent of $\mu$, and each choice of $p$ describes a point on the zoo plot for fixed $N$. The uncertainty in $N$ turns each of these points into lines, which smear together into a continuous region in Fig. \ref{fig:modelslog}. Note that even if we include all of these uncertainties, the different classes of potential do not have significant overlap on the zoo plot, and it is therefore possible to distinguish one from another observationally. Furthermore, for a given choice of potential, the uncertainties in $r$ and $n$ arising from the uncertainty in $N$ are generally strongly correlated. This correlation will be apparent in the flow analysis presented below. Finally, for particular choices of potential such as the exponential potential, inflation formally continues forever and the uncertainty due to $N$ vanishes altogether, so there is no ``smearing'' of the line on the zoo plot.

\section{Flow Equations}
\label{sec:flowequations}

In this section we describe the flow equations which are a useful
tool to quantify the theoretical errors in the inflationary 
observables due to our ignorance about of the number of {\it e}-foldings.

We have defined the slow roll parameters $\epsilon$ and $\eta$ in terms of
the Hubble parameter $H\left(\phi\right)$ as
\begin{eqnarray}
\epsilon &\equiv& {m_{\rm Pl}^2 \over 4 \pi} \left({H'(\phi) \over
 H(\phi)}\right)^2,\cr
\eta\left(\phi\right) &\equiv& {m_{\rm Pl}^2 \over 4 \pi} 
\left({H''\left(\phi\right)
\over H\left(\phi\right)}\right).
\end{eqnarray}
These parameters are simply related to observables $r \simeq 16 \epsilon$, and
$n - 1 \simeq 4 \epsilon - 2 \eta$ to first order in slow roll.
 Taking higher derivatives
of $H$ with respect to the field, we can define an infinite hierarchy of slow
roll parameters \cite{liddle94}:
\begin{eqnarray}
\sigma &\equiv& {m_{\rm Pl} \over \pi} \left[{1 \over 2} \left({H'' \over
 H}\right) -
\left({H' \over H}\right)^2\right],\cr
{}^\ell\lambda_{\rm H} &\equiv& \left({m_{\rm Pl}^2 \over 4 \pi}\right)^\ell
{\left(H'\right)^{\ell-1} \over H^\ell} {d^{(\ell+1)} H \over d\phi^{(\ell +
1)}}.
\end{eqnarray}
Here we have chosen the parameter $\sigma \equiv 2 \eta - 4 \epsilon \simeq n
-1 $ to make comparison with observation convenient.

For our purposes, it is convenient to use $N$ as the measure of time during inflation. As above,
we take $t_e$ and $\phi_e$ to be the time and field value at end of
inflation. Therefore, $N$ is defined as the number of {\it e}-folds before the end of
inflation, and increases as one goes {\em backward} in time ($d t > 0
\Rightarrow d N < 0$):
\begin{equation}
{d \over d N} = {d \over d\ln a} = { m_{\rm Pl} \over 2 \sqrt{\pi}}
\sqrt{\epsilon} {d \over d\phi},
\end{equation}
where we have chosen the sign convention that $\sqrt{\epsilon}$ has the same
sign as $H'\left(\phi\right)$:
\begin{equation}
\sqrt{\epsilon} \equiv + {m_{\rm PL} \over 2 \sqrt{\pi}} {H' \over H}.
\end{equation}
Then $\epsilon$ itself can be expressed in terms of $H$ and $N$ simply as,
\begin{equation}
\label{eqepsilonfromN}
{1 \over H} {d H \over d N} = \epsilon.
\end{equation}
Similarly, the evolution of the higher order parameters during inflation is
determined by a set of ``flow'' equations \cite{hoffman,sc,kinneyflow},
\begin{eqnarray}
{d \epsilon \over d N} &=& \epsilon \left(\sigma + 2
\epsilon\right),\cr {d \sigma \over d N} &=& - 5 \epsilon \sigma - 12
\epsilon^2 + 2 \left({}^2\lambda_{\rm H}\right),\cr {d
\left({}^\ell\lambda_{\rm H}\right) \over d N} &=& \left[
\frac{\ell - 1}{2} \sigma + \left(\ell - 2\right) \epsilon\right]
\left({}^\ell\lambda_{\rm H}\right) + {}^{\ell+1}\lambda_{\rm
H}.\label{eqfullflowequations}
\end{eqnarray}
The derivative of a slow roll parameter at a given order is higher order in
slow roll. At the lowest order, this set of equations properly expressed
in terms of observables reproduce equations (\ref{flow1}).

A boundary condition can be specified at any point in the
inflationary evolution by selecting a set of parameters
$\epsilon,\sigma,{}^2\lambda_{\rm H},\ldots$ for a given value of $N$. This is
sufficient to specify a ``path'' in the inflationary parameter space that
specifies the evolution of the observables in terms of the number of {\it e}-foldings. 
Taken to infinite order,
this set of equations completely specifies how a shift in the
number of {\it e}-foldings is reflected in a shift of the
slow roll parameters and, therefore, of the observables.  
Furthermore, such a
quantification  is exact, 
with no assumption of slow roll necessary. In practice,
we must truncate the expansion at finite order by assuming that the
${}^\ell\lambda_{\rm H}$ are all zero above some fixed value of $\ell$.  

Once we obtain a solution to the flow equations
$[\epsilon(N),\sigma(N),{}^\ell\lambda_{\rm H}(N)]$, we can calculate the
predicted values of the tensor/scalar ratio $r$, the spectral index $n$, and
the ``running'' of the spectral index $d n / d\ln k$ and how
they change upon shifting the number of {\it e}-foldings by $\Delta N$.  
To lowest order, the
relationship between the slow roll parameters and the observables is especially
simple: $r = 16 \epsilon$, $n - 1 = \sigma$, and $d n / d \ln k = 0$. To
second order in slow roll, the observables are given by
\cite{liddle94,stewart93},
\begin{equation}
r = 16 \epsilon \left[1 - C \left(\sigma + 2
 \epsilon\right)\right],\label{eqrsecondorder}
\end{equation}
for the tensor/scalar ratio, and 
\begin{equation}
n - 1 = \sigma - \left(5 - 3 C\right) \epsilon^2 - {1 \over 4} \left(3
- 5 C\right) \sigma \epsilon + {1 \over 2}\left(3 - C\right)
\left({}^2\lambda_{\rm H}\right)\label{eqnsecondorder}
\end{equation}
for the spectral index. The constant $C \equiv 4 (\ln{2} +
\gamma) - 5 = 0.0814514$, where $\gamma \simeq 0.577$ is Euler's
constant.
 Derivatives
with respect to wavenumber $k$ can be expressed in terms of derivatives with
respect to $N$ as \cite{liddle95}
\begin{equation}
{d \over d N} = - \left(1 - \epsilon\right) {d \over d \ln k},
\end{equation}
The scale dependence of $n$ is then  given by the simple expression
\begin{equation}
{d n \over d \ln k} = - \left({1 \over 1 - \epsilon}\right) {d n \over d N},
\end{equation}
which can be evaluated by using Eq.~(\ref{eqnsecondorder}) and the flow
equations.  

\begin{figure}
\includegraphics[width=3.1in]{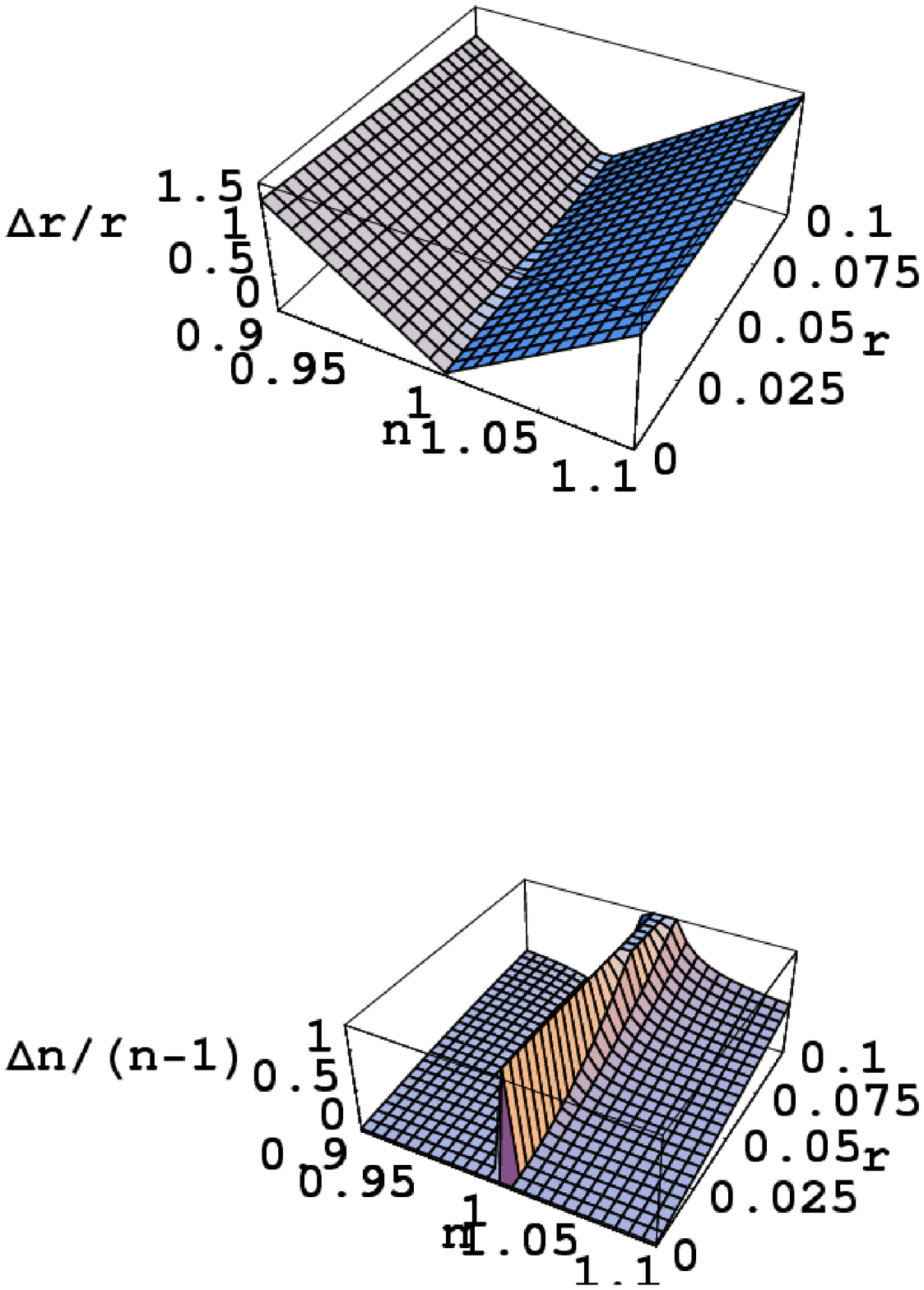}
\caption{\label{fig:lowestordererrors} Lowest order estimates (\ref{eq:lowestordererrors}) of the theoretical errors $\Delta r / r$ (top) and $\Delta n$ (bottom) as functions of $n$ and $r$, assuming $\Delta N = 14$.}
\end{figure}

It is straightforward to use the flow equations to obtain lowest-order estimates of the expected theoretical errors $\Delta r$ and $\Delta n$ in the predictions for $r$ and $n$ by adopting the simple approximation
\begin{eqnarray}
\Delta r &&\sim {d r \over d N} \Delta N\cr
\Delta n &&\sim {d n \over d N} \Delta N,
\end{eqnarray}
where
\begin{eqnarray}
{d r \over d N} \simeq 16 {d \epsilon \over d N} &&= 16 \epsilon \left(\sigma + 2 \epsilon\right))\cr
&&= r \left[\left(n - 1\right) + \left(r / 8\right)\right],
\end{eqnarray}
and
\begin{eqnarray}
{d n \over d N} \simeq {d \sigma \over d N} &&= -5 \epsilon \sigma - 12 \epsilon^2 + 2 \left({}^2\lambda_{\rm H}\right)\cr
&&= - {5 \over 16} r \left(n - 1\right) - {3 \over 32} r^2 + 2 \left({}^2\lambda_{\rm H}\right).
\end{eqnarray}
We then have estimates for the uncertainties $\Delta r$ and $\Delta n$ in terms of the uncertainty in the number of {\it e}-folds $\Delta N$:
\begin{eqnarray}
\label{eq:lowestordererrors}
{\Delta r \over r} &&\sim \left[\left(n - 1\right) + \left(r / 8\right)\right] \left(\Delta N\right)\cr
{\Delta n \over n - 1} &&\sim \left[{5 \over 16} r + {3 \over 32} {r^2 \over n - 1}\right] \left(\Delta N\right),
\end{eqnarray}
where we have taken ${}^2\lambda_{\rm H} \simeq 0$. Figure \ref{fig:lowestordererrors} shows the error estimates from Eq. (\ref{eq:lowestordererrors}) as a function of $r$ and $n$. 
These estimates indicate that the theoretical errors in $n$ and $r$ can be substantial, depending on where the model lives in the $r$-$n$ plane. If we take the region roughly favored by current observation, $r < 0.1$ and $\left\vert n - 1\right\vert < 0.1$, we have upper bounds on the errors of order 
\begin{eqnarray}
{\Delta r \over r} &&\leq 1\cr
{\Delta n \over \left\vert n - 1\right\vert} &&\leq 1,
\end{eqnarray}
where these bounds are saturated for $r \sim 0.1$, $\left\vert n - 1\right\vert \sim 0.1$. Note in particular that, for $n - 1 \sim 0.1$, the fractional error in $r$ is largely {\em independent} of the value of r.
The fractional errors $\Delta r / r$ and $\Delta n / \left\vert n - 1\right\vert$ are first order in slow roll. The absolute error $\Delta n$ is second order in slow roll, and since 
\begin{equation}
\Delta n \propto {d n \over d \ln{k}},
\end{equation}
we expect substantial absolute error in the spectral index only in models which also predict a relatively large running $d n / d \ln{k}$. These estimates indicate that the theoretical errors in the inflationary observables can be significant compared to the expected accuracy of future observational constraints. Note that for a single-parameter set of models such as the large-field case (\ref{large}), there exists an $N$-independent relation between $r$ and $n$. Therefore the errors in the parameters are highly correlated. Such models, as we have seen, are falsifiable by observation. Such a simple relation between the observables will not exist for models described by a larger number of parameters. In the next section, we present the results of a Monte Carlo analysis which extends these estimates to higher order in slow roll, effectively increasing the dimensionality of the parameter space describing the potentials. 

\section{Monte Carlo estimate of theoretical errors}
\label{sec:numericalresults}

In Sec. \ref{sec:flowequations} we derived an analytical estimate of the theoretical errors in the observables $n$ and $r$ to lowest order in slow roll of
\begin{eqnarray}
{\Delta r \over r} &&\leq 1\cr
{\Delta n \over \left\vert n - 1\right\vert} &&\leq 1
\end{eqnarray}
While higher-order analogs of Eq. (\ref{eq:lowestordererrors}) are in principle possible to derive using the flow equations, a comprehensive investigation of the effect of higher-order terms in slow roll is best accomplished using numerical techniques. In this section, we discuss the results of using a Monte Carlo evaluation of the flow equations to determine the errors $\Delta r / r$ and $\Delta n / \left\vert n - 1\right\vert$ for a large ensemble of inflationary models. 

Monte Carlo evaluation of the flow equations, introduced in Ref. \cite{kinneyflow}, has become a standard technique for investigating the inflationary model space. The principle is straightforward: since the flow equations (\ref{eqfullflowequations}) are first order differential equations, the selection of a point in the slow roll parameter space $\left\lbrace \epsilon, \eta, {}^2\lambda_{\rm H}, \ldots\right\rbrace$ serves to completely specify the evolution of a particular model in the space of slow roll parameters. For a model specified in this way, there is a straightforward  procedure for determining its observable predictions, that is, the values of  $r$, $n - 1$, and $d n / d \ln k$ a fixed number $N$ {\it e}-folds before the end of inflation. The algorithm for a single model is as follows:
\begin{itemize}
\item{Select a point in the parameter space $\epsilon,\eta,{}^l\lambda_{\rm
 H}$.}
\item{Evolve forward in time ($d N < 0$) until either (a) inflation ends, or (b)
 the evolution reaches a late-time fixed point.}
\item{If the evolution reaches a late-time fixed point, calculate the
 observables $r$, $n - 1$, and $d n / d \ln k$ at this point.}
\item{If inflation ends, evaluate the flow equations backward $N$ {\it e}-folds from
 the end of inflation. Calculate the observable parameters at this point.}
\end{itemize}
The end of inflation is given by the condition $\epsilon = 1$. In principle, it is possible to carry out this program exactly, with no assumptions made about the convergence of the hierarchy of slow roll parameters. In practice, the series of flow equations  (\ref{eqfullflowequations}) must be truncated at some finite order and evaluated numerically. The calculations presented here are performed to eighth order in slow roll. \footnote{The reader is referred to Ref. \cite{kinneyflow} for a more detailed discussion of the procedure used to stochastically evaluate the flow equations.} In effect, we are expanding our model space from the set of single-parameter potentials considered in Sec. \ref{seczoology} to consider potentials with eight free parameters describing their shape \cite{Liddle:2003py}. 

We wish to determine how the uncertainty in the total number of {\it e}-folds $N$ translates into a uncertainties in the observable parameters $r$ and $n$. For models which reach a late-time attractor $r = 0$, $n > 1$ in the flow space, the answer is trivial: since the observables are evaluated at a fixed point in the flow space, the shift in the observables with $N$ by definition vanishes, and the theoretical uncertainty is in this sense negligible.\footnote{A more realistic hybrid-type model displaying this dynamics will likely be more complex, since the field may not yet have settled into the attractor solution at the appropriate point for calculating cosmological observables. Such a situation is highly model-dependent, and we do not consider it further here.} We therefore concentrate on models dubbed ``nontrivial'' in the language of Ref. \cite{kinneyflow}, that is models for which the dynamics carry the evolution through $\epsilon = 1$ and inflation naturally ends after a finite number of {\it e}-folds. For a given solution to the flow equations, it is simple to evaluate the effect of moving along the ``path'' in flow space from $N = 46$ to $N = 60$. (Ref. \cite{Chongchitnan:2005pf} contains an interesting analytic analysis of the dynamics of paths in the space of flow parameters.)

\begin{figure}
\includegraphics[width=3.25in]{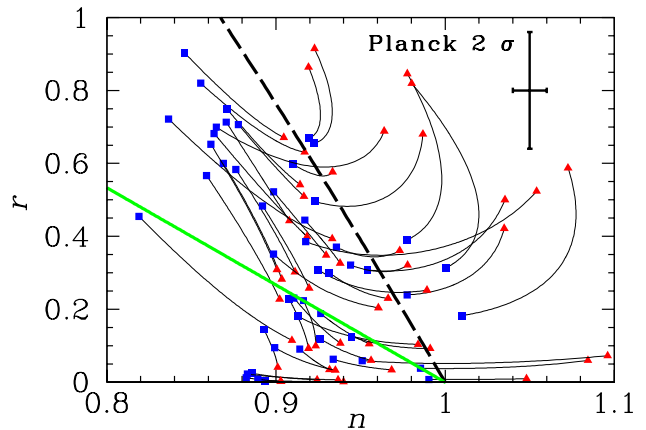}
\caption{\label{fig:nr} Flow from $N = 46$ (blue squares) to $N = 60$ (red triangles) for an ensemble of fifty models generated via flow Monte Carlo, plotted in the $n$ - $r$ plane. The path traced by the flow indicates the level of theoretical uncertainty induced by the uncertainty $\Delta N$. The diagonal lines indicate the boundares between small-field and large-field (green, solid) and large-field and hybrid (black, dashed). Models can ``shift'' class from $N = 46$ to $N = 60$. The error bar at top right shows projected $2\sigma$ measurement uncertainties in $n$ and $r$ for the Planck satellite. The central value for the error bar shown is arbitrary.}
\end{figure}

\begin{figure}
\includegraphics[width=3.25in]{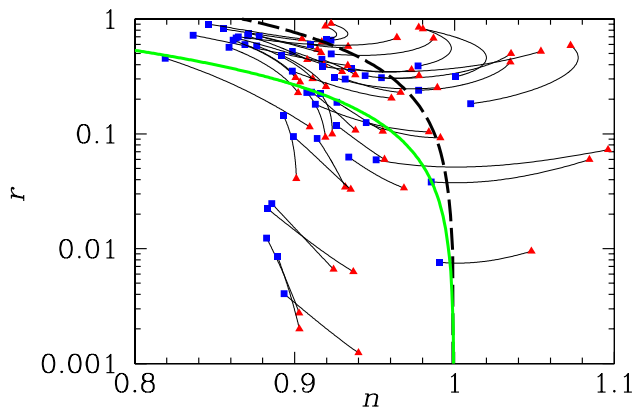}
\caption{\label{fig:nrlog} Flow from $N = 46$ (blue squares) to $N = 60$ (red triangles) for an ensemble of fifty models generated via flow Monte Carlo, plotted in the $n$ - $\log{r}$ plane, showing the behavior for small $r$. }
\end{figure}

\begin{figure}
\includegraphics[width=3.25in]{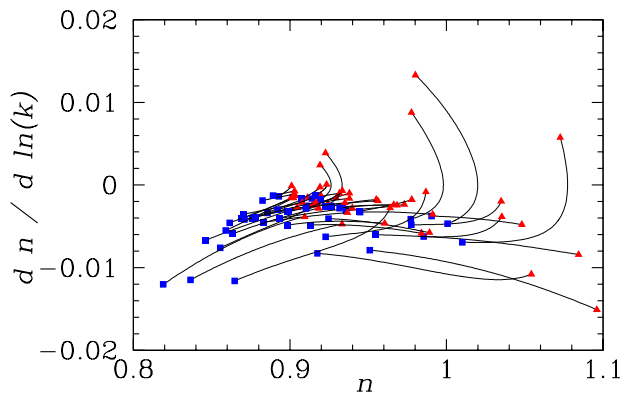}
\caption{\label{fig:nnk} Flow from $N = 46$ (blue squares) to $N = 60$ (red triangles) for an ensemble of fifty models generated via flow Monte Carlo, plotted in the $n$ - $dn / d \ln{k}$ plane. }
\end{figure}

Figures \ref{fig:nr}, \ref{fig:nrlog} and \ref{fig:nnk} 
illustrate the effect of shifting the number of {\it e}-folds in the space of the 
observables $r$, $n$, and $d n / d \ln{k}$. We see that the effect of the theoretical 
error in $N$ on the values of the observables is substantial, and at least qualitatively 
consistent with our rough estimate (\ref{eq:lowestordererrors}) and with our previous 
discussion for large and small field models. 
For large field models and for moderate
value of $(1-n)$ and $r$, for which the first order approximations hold, we see that
the errors increase respectively with $(1-n)$ and $r$. Moving towards
larger values of these parameters implies a substantial role played by higher order
corrections and errors can be very sizable. For small field models, we observe 
large displacements along the $(n-1)$-axis, but small
ones along the $r$-axis.

We can also see from Fig. \ref{fig:nr} that the 
lowest-order classification of models into small field, large field, 
and hybrid breaks down when higher order corrections to the dynamics are included. 
Models which fall into the large-field region at $N = 46$ can evolve into the small-field
 region at $N = 60$, a behavior which was noted in Ref. \cite{Schwarz:2004tz}.

\begin{figure}
\includegraphics[width=3.25in]{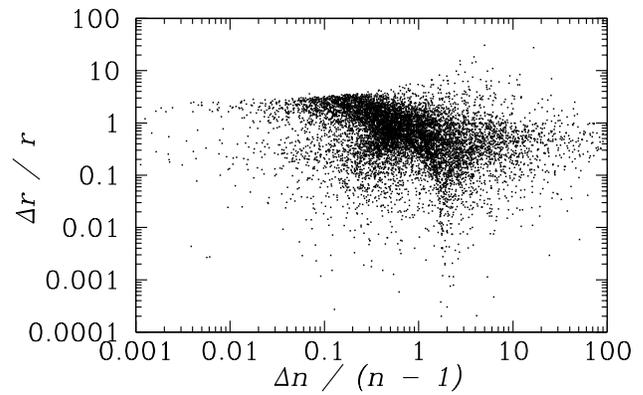}
\caption{\label{fig:delndelr} Theoretical uncertainties $\Delta n$ and $\Delta r / r$ calculated for an ensemble of 10,000 models with $0.9 < n_{60} < 1.1.$ and $r_{60} > 0.001$. The models cluster strongly in the region $\Delta r / r \sim 0.1 - 1$ and $\Delta n / \left\vert n - 1\right\vert \sim 0.1 - 1$, consistent with the lowest-order estimate (\ref{eq:lowestordererrors}). }
\end{figure}

\begin{figure}
\includegraphics[width=3.25in]{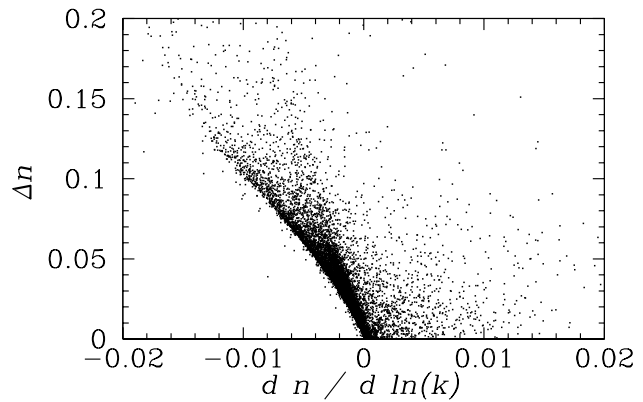}
\caption{\label{fig:nkdeln} The absolute uncertainty in the spectral index $\Delta n$ plotted versus the running $d n / d \ln{k}$. As expected, the uncertainty in the spectral index becomes large for models which predict significant running.}
\end{figure}

To obtain a more quantitative understanding of the theoretical error in the observables induced by the uncertainty $\Delta N$, we generate an ensemble of models using the flow equations and calculate the observables at $N= 60$, denoted $r_{60}$ and $n_{60}$, and at $N = 46$, denoted $r_{46}$ and $n_{46}$. We retain only models which lie close to the region observationally favored by WMAP, $0.9 < n_{60} < 1.1$, and we retain only models with a non-negligible tensor amplitude, $r_{60} > 0.001$. For each model generated by the Monte Carlo, we then assign uncertainties in $r$ and $n$ as:
\begin{equation}
{\Delta r \over r} = \left\vert{r_{60} - r_{46} \over r_{60}}\right\vert,
\end{equation}
and
\begin{equation}
{\Delta n \over \left\vert n - 1\right\vert} = \left\vert{n_{60} - n_{46} \over n_{60} - 1}\right\vert.
\end{equation}
Figure \ref{fig:delndelr} shows the above uncertainties calculated for an ensemble of 10,000 models. We see that the estimates $\Delta r / r \sim \Delta n / \left\vert n - 1\right\vert \sim 1$ are robust even when calculated to higher order. Figure \ref{fig:nkdeln} shows the uncertainty $\Delta n$ plotted against the running of the spectral index $dn / d \ln{k}$, showing the expected strong correlation between the error $\Delta n$ and the running. 

\section{Conclusions}

We have considered the implications of the uncertainty in the reheat temperature on the observable predictions of inflation. The most convenient parameters for discriminating among inflation models are the tensor/scalar ratio $r$ and the scalar spectral index $n$. These parameters are simply related to the inflationary slow roll parameters $\epsilon$ and $\eta$, and this correspondence can be generalized to higher order in the slow roll expansion through the inflationary flow equations. The uncertainty in the reheat temperature corresponds to uncertainty in the number of {\it e}-folds of expansion $N$ during the inflationary epoch, 
\begin{equation}
N\simeq 60+\frac{1}{6}{\rm ln}\,\left(-n_T\right)+\frac{1}{3}\,{\rm ln}
\left(T_{\rm RH}/10^{16}\,{\rm GeV}\right)-\frac{1}{3}\,{\rm ln}\,\gamma\, ,
\end{equation}
where $T_{\rm RH}$ is the reheating temperature, $\gamma$ is the ratio of the
entropy per comoving volume today to that after reheating. If we assume $\gamma$ is negligible, the large uncertainty in the reheat temperature corresponds to an uncertainty in the number of {\it e}-folds $\Delta N \sim 14$. This can be related to a theoretical uncertainty in the observable parameters $r$ and $n$ to lowest order by using the flow relations
\begin{eqnarray}
{\Delta r \over r} &&\sim \left[\left(n - 1\right) + \left(r / 8\right)\right] \left(\Delta N\right),\cr
{\Delta n \over n - 1} &&\sim \left[{5 \over 16} r + {3 \over 32} {r^2 \over n - 1}\right] \left(\Delta N\right). 
\end{eqnarray}
For $r$ and $n$ in the region favored by current observational constraints, these errors can in principle be large, $\Delta r / r \sim 0.1 - 1$, and $\Delta n / \left\vert n - 1\right\vert \sim 0.1 - 1$.  
We also analyze the expected theoretical uncertainty in the inflationary observables by Monte Carlo evaluation of the inflationary flow equations to eighth order in slow roll, and find results consistent with the lowest-order estimate above, but with considerable scatter in the models. We have numerically checked that these  errors increase approximately linearly with $\Delta N$ as suggested by the lowest-order estimate. The absolute error in the spectral index can be compared with the expected $2\sigma$ uncertainty in the spectral index from the Planck satellite, $\Delta n \sim 0.01$, and is seen to be typically of the same order.

We conclude that the theoretical uncertainties in the inflationary observables are generically of the order of or larger than the projected uncertainties in future precision measurements of the Cosmic Microwave Background, and represent a significant challenge for the program of using observation to distinguish among the many different candidate models for inflation in the early universe. While the dependence of the inflationary observables on the number of {\it e}-folds $N$ is certainly well known (and was, for example, taken into account by WHK in Refs. \cite{kinney98a,kinneyflow}), it has not been emphasized as the dominant source of theoretical error in the predictions of inflation. The error induced by the 
uncertainty in the reheat temperature and/or in the amount of entropy release
after inflation (and thus in $N$) is typically much larger than errors in the quantities $r$ and $n$ due to using the slow roll approximation to calculate the primordial power spectrum, a subject which has received considerable attention in the literature \cite{sc,Gong:2001he,Habib:2002yi,Habib:2004kc,Casadio:2004ru,Makarov:2005uh}.  The expected uncertainties for any {\em particular} choice of potential are model-dependent: it will certainly still be possible to rule out models of inflation with future precision data. However, it will be necessary to move beyond the simple lowest-order description of the inflationary parameter space which has so far been good enough.

\section*{Acknowledgments}
AR is on leave of absence from INFN Padova, Italy. We thank Brian Powell for work on an upgraded flow code. WHK is supported in part by the National Science Foundation under grant NSF-PHY-0456777.




\begin{thebibliography}{99}
\frenchspacing


\bibitem{lrreview} For reviews, see  
D. H. Lyth and A. Riotto, 
Phys. Rept. {\bf 314}, 1 (1999); W.~H.~Kinney,
  arXiv:astro-ph/0301448.

\bibitem{Starobinsky:1979ty}
  A.~A.~Starobinsky,
  JETP Lett.\  {\bf 30}, 682 (1979)
  [Pisma Zh.\ Eksp.\ Teor.\ Fiz.\  {\bf 30}, 719 (1979)].

\bibitem{muk81} 
V. F. Mukhanov and G. V. Chibisov, 
JETP  Lett. {\bf 33}, 532  (1981).

\bibitem{bardeen83} 
J. M. Bardeen, P. J. Steinhardt, and M. S. Turner, 
Phys. Rev. D {\bf 28}, 679 (1983).

\bibitem{bennett96} 
C. L. Bennett 
{\it et al.} Astrophys. J. {\bf 464}, L1 (1996).

\bibitem{gorski96} 
K. M. Gorski 
{\it et al.} Astrophys. J. {\bf 464}, L11 (1996).

\bibitem{wmap1} 
C. L. Bennett {\it et al.},
Astrophys. J. Suppl. {\bf 148}, 1 (2003).


\bibitem{ex} 
H. V. Peiris {\it et al.},
Astrophys. J. Suppl.  {\bf 148}, 213 (2003);
W. H. Kinney, E. W. Kolb, A. Melchiorri and A. Riotto,
Phys. Rev. D {\bf 69}, 103516 (2004).


\bibitem{planck} See {\tt  http://www.rssd.esa.int/index.php?project=PLANCK}.
\bibitem{CMBpol} See {\tt 
www.mssl.ucl.ac.uk/www${}_{-}$astro/submm/CMBpol1.html}.
\bibitem{south} See {\tt http://astro.uchicago.edu/spt/}.


\bibitem{knox} M.~Kaplinghat, L.~Knox and Y.~S.~Song,
  Phys.\ Rev.\ Lett.\  {\bf 91}, 241301 (2003).

\bibitem{gravex} 
M. Kesden, A. Cooray and M. Kamionkowski, 
Phys. Rev. Lett. {\bf 89} (2002) 011304; 
L. Knox and Y.-Song, 
Phys. Rev. Lett. {\bf 89} (2002) 011303.

\bibitem{BBO}
S. Phinney, et al., ``The Big Bang Observer: Direct detection of
gravitational waves from the birth of the universe to the present,''
NASA mission concept study (2005).

\bibitem{decarlos}  B.~de Carlos, J.~A.~Casas, F.~Quevedo and E.~Roulet,
  Phys.\ Lett.\ B {\bf 318}, 447 (1993).

\bibitem{ll} A.~R.~Liddle and S.~M.~Leach,
  Phys.\ Rev.\ D {\bf 68}, 103503 (2003).

\bibitem{kinneyflow} 
W.~H.~Kinney, Phys. Rev. D {\bf 66}, 083508 (2002).

\bibitem{hoffman}
M.~B.~Hoffman and M.~S.~Turner, Phys. Rev. D {\bf 64}, 023506 (2001).

\bibitem{sc} D.~J.~Schwarz, C.~A.~Terrero-Escalante, and A.~.A.~Garcia,
Phys. Lett. {\bf B517}, 243 (2001).


\bibitem{grishchuk88} 
L. P. Grishchuk and Yu. V. Sidorav, in {\it Fourth Seminar on Quantum Gravity},
eds M. A. Markov, V. A. Berezin and V. P. Frolov (World Scientific, Singapore,
1988).

\bibitem{muslimov90} 
A. G. Muslimov, Class. Quant. Grav. {\bf 7}, 231 (1990).

\bibitem{salopek90} 
D. S. Salopek and J. R. Bond, Phys. Rev. D {\bf 42}, 3936 (1990).

\bibitem{lidsey95} 
J. E. Lidsey {\it et al.}, Rev. Mod. Phys. {\bf 69}, 373 (1997),
astro-ph/9508078.



\bibitem{dodelson97} 
S.~Dodelson, W.~H.~Kinney, and E.~W.~Kolb, Phys. Rev. D {\bf 56}, 3207 (1997), 
astro-ph/9702166.

\bibitem{kinney98a} 
W.~H.~Kinney, Phys. Rev. D {\bf 58}, 123506 (1998).


\bibitem{linde83} 
A. D. Linde, Phys. Lett. {\bf 129B}, 177 (1983).




\bibitem{linde82} 
A.~D.~Linde, Phys.\ Lett.\  {\bf B108} 389, 1982.

\bibitem{albrecht82} 
A. Albrecht and P. J. Steinhardt, Phys. Rev. Lett {\bf48}, 1220 (1982).



\bibitem{freese90} 
K. Freese, J. Frieman, and A. Olinto, Phys. Rev. Lett {\bf 65}, 3233 (1990).


\bibitem{linde91} 
A. D. Linde, Phys. Lett. {\bf 259B}, 38 (1991).

\bibitem{linde94} 
A. D. Linde, Phys. Rev. D {\bf 49}, 748 (1994).

\bibitem{copeland94} 
E. J. Copeland, A. R. Liddle, D. H. Lyth, E. D. Stewart, and D. Wands,
Phys. Rev. D {\bf 49}, 6410 (1994); 
A.~D.~Linde and A.~Riotto,
  Phys.\ Rev.\ D {\bf 56}, 1841 (1997).


\bibitem{barrow93} 
J. D. Barrow and A. R. Liddle, Phys. Rev. D {\bf 47}, R5219 (1993).


\bibitem{kinney97} 
W. H. Kinney and A. Riotto, Astropart. Phys. {\bf 10}, 387 (1999).


\bibitem{kinney98} 
W. H. Kinney and A. Riotto, Phys. Lett.{\bf 435B}, 272 (1998).

\bibitem{liddle94} 
A.~R.~Liddle, P. Parsons, and J. D. Barrow, Phys. Rev. D {\bf 50}, 7222 (1994).

\bibitem{stewart93} 
E. D. Stewart and D. H. Lyth, Phys. Lett. {\bf 302B}, 171 (1993).

\bibitem{liddle95} 
A.~R.~Liddle. and M.~S.~Turner, Phys. Rev. D {\bf 50}, 758 (1994). 

\bibitem{Liddle:2003py}
  A.~R.~Liddle,
  Phys.\ Rev.\ D {\bf 68}, 103504 (2003)
  [arXiv:astro-ph/0307286].

\bibitem{Chongchitnan:2005pf}
  S.~Chongchitnan and G.~Efstathiou,
  arXiv:astro-ph/0508355.

\bibitem{Schwarz:2004tz}
  D.~J.~Schwarz and C.~A.~Terrero-Escalante,
  JCAP {\bf 0408}, 003 (2004)
  [arXiv:hep-ph/0403129].

\bibitem{Gong:2001he}
  J.~O.~Gong and E.~D.~Stewart,
  Phys.\ Lett.\ B {\bf 510}, 1 (2001)
  [arXiv:astro-ph/0101225].

\bibitem{Habib:2002yi}
  S.~Habib, K.~Heitmann, G.~Jungman and C.~Molina-Paris,
  Phys.\ Rev.\ Lett.\  {\bf 89}, 281301 (2002)
  [arXiv:astro-ph/0208443].

\bibitem{Habib:2004kc}
  S.~Habib, A.~Heinen, K.~Heitmann, G.~Jungman and C.~Molina-Paris,
  Phys.\ Rev.\ D {\bf 70}, 083507 (2004)
  [arXiv:astro-ph/0406134].

\bibitem{Casadio:2004ru}
  R.~Casadio, F.~Finelli, M.~Luzzi and G.~Venturi,
  Phys.\ Rev.\ D {\bf 71}, 043517 (2005)
  [arXiv:gr-qc/0410092].

\bibitem{Makarov:2005uh}
  A.~Makarov,
  arXiv:astro-ph/0506326.


\end{thebibliography}
\end{document}